\documentclass[final,1p,times]{elsarticle}

\usepackage{amssymb}
\usepackage{amsthm}
\usepackage{amsmath}
\usepackage{listings} 
\usepackage{subcaption}
\usepackage{graphicx}

\usepackage{xcolor} 

\journal{Applied Ergonomics}

\begin{document}

\begin{frontmatter}

\title{
Just Roll with It: Exploring the Mitigating Effects of Postural Alignment on Vection-Induced Cybersickness in Virtual Reality Over Time.
}

\author[inst1]{Charlotte Croucher}

\affiliation[inst1]{organization= {Department of Cognitive Robotics, Delft University of Technology},
            city={Delft},
            postcode={2628 CD}, 
            country={The Netherlands}}

\author[inst2,inst3,inst4]{Panagiotis Kourtesis}

\affiliation[inst2]{organization={Department of Psychology, The American College of Greece},
            city={Athens},
            postcode={15342}, 
            country={Greece}
            }
            
\affiliation[inst3]{organization={Department of Informatics and Telecommunications, National and Kapodistrian University of Athens},
            city={Athens},
            postcode={16122}, 
            country={Greece}
            }
\affiliation[inst4]{organization={Department of Psychology, The University of Edinburgh},
            city={Edinburgh},
            postcode={EH8 9Y}, 
            country={United Kingdom}
            }

\author[inst1]{Georgios Papaioannou}

\begin{abstract}
Cybersickness remains a significant challenge in virtual reality (VR), limiting its usability across various applications. Existing mitigation strategies focus on optimising VR hardware and/or software and enhancing self-motion perception to minimise sensory conflict. 
However, anticipatory postural adaptation, a strategy widely studied with regards to motion sickness while being driven, has not been systematically examined in VR. 
Therefore, in this study, we explore whether adopting comfort-orientated postural movements, based on the literature, mitigates cybersickness. 
We conducted an exploratory analysis using a cumulative link mixed model (CLMM) on secondary data from a VR-based postural alignment experiment. 
Results indicate that misalignment between trunk roll and the virtual trajectory increases the odds of reporting higher cybersickness scores by 5\%. Additionally, each additional minute in VR increases the odds of reporting higher cybersickness scores (FMS scores) by 11\%, but prolonged exposure leads to a 75\% reduction in the odds of reporting cybersickness symptoms, suggesting adaptation effects. 
Individual differences also play a role, with higher cybersickness susceptibility increasing the odds of reporting higher symptom severity by 8\%. These findings indicate that anticipatory postural adaptation could serve as a natural mitigation strategy for cybersickness. 
VR applications, particularly in training and simulation, may benefit from designing adaptive cues that encourage users to align their posture with virtual movement. 
Future research should explore real-time postural feedback mechanisms to enhance user comfort and reduce cybersickness.

\end{abstract}


\begin{highlights}
\item Misalignment of trunk roll increases the odds of reporting higher cybersickness by 5\%
\item Each minute in VR increases the odds of reporting higher cybersickness by 11\%
\item Prolonged VR exposure decreases the odds of reporting higher cybersickness by 75\%.
\item Higher cybersickness susceptibility increases the odds of reporting higher scores by 8\%
\end{highlights}


\end{frontmatter}


\section{Introduction}
\label{Intro}

Virtual Reality (VR) is a medium that immerses the users into 3D virtual environments, allowing a simulation of the real world, yet without its physical constraints \cite{SlaterSanchez2016, Kourtesis2024b}. Contemporary VR systems render virtual environments on head-mounted displays (HMDs) that facilitate deeper immersion (i.e., a sense of presence and embodiment) and intuitive ergonomic interactions (i.e., mimicking real-world interactions) \cite{Kourtesis2024b}. For these reasons, VR has been used for various purposes, such as immersive learning in education, medical and professional training, neuropsychological evaluation, therapeutic interventions, arts, entertainment, retail, e-commerce, remote work, sports, architecture, urban planning and preservation of cultural heritage \cite{SlaterSanchez2016, Kourtesis2024b}.

Many applications of VR require virtual motion, known as vection (i.e., the illusion of self-motion caused by visual stimuli in the absence of corresponding physical movement) \cite{Stanney2020, Davis2014}. For entertainment purposes, a user could experience a virtual roller coaster in the comfort of their own home or even ride an actual roller coaster while wearing a VR HMD \cite{Burt2019}. For training-based scenarios, applications may include preflight training for astronauts \cite{Chen2015} and fall prevention training for rehabilitation patients \cite{Phu2019}. However, despite these benefits and extensive application areas, a pervasive issue that hinders the widespread adoption of VR is cybersickness \cite{Kourtesis2024b}.

Cybersickness manifests as nausea, dizziness, disorientation, and oculomotor disturbances \cite{MazloumiGavgani2018}, with symptoms varying in severity across individuals \cite{MazloumiGavgani2018}. The most widely accepted theoretical explanations for cybersickness include the sensory conflict theory, which posits that symptoms arise when visual and vestibular cues are misaligned \cite{Reason1978}. Another theoretical explanation is the postural instability theory, which suggests that instability in postural control exacerbates cybersickness symptoms \cite{Riccio1991, Stoffregen1998}. Given that vection is one of the primary triggers of cybersickness in VR and a key driver of this sensory mismatch, mitigating vection-induced cybersickness is essential to improving VR usability across various domains \cite{LaViola2000, Davis2014}.

A wide range of cybersickness mitigation strategies have been explored, many of which target vection or sensory conflict directly. Hardware and software optimisations, such as reducing latency through improved rendering algorithms, aimed to decrease the delay between visual motion cues and vestibular feedback \cite{Kundu2021}. Sensory augmentation techniques, such as vibrotactile stimulation in the torso, seating, or floor, have improved self-motion perception and reduced sensory conflicts \cite{Phlmann2024, Jung2021, Grassini2021}. Furthermore, aligning visual and olfactory cues - by introducing pleasant odours that match the virtual environment - has been shown to moderate vection-induced discomfort \cite{Reichl2024}. Visual field manipulations, such as applying motion blur, subtle geometric distortions, and opaque occlusion, also aim to alter vection perception and mitigate cybersickness symptoms \cite{Groth2021, Groth2024}. 
Furthermore, galvanic vestibular stimulation has been explored to provide congruent vestibular feedback and reduce vection-induced cybersickness \cite{Weech2020}. However, these approaches focus on modifying sensory input rather than how individuals physically respond to vection. The role of postural alignment in mitigating vection-induced cybersickness remains largely unexplored despite its potential to influence sensory integration and motion perception.

Given that vection-induced cybersickness closely resembles motion sickness (MS) \cite{Lukacova2023, Keshavarz2023}, research on MS mitigation may offer valuable insights into reducing vection-induced discomfort. From this perspective, vehicle design optimisations, such as improved motion control and seat stabilisation, help minimise abrupt accelerations and whole-body vibrations, thereby reducing MS symptoms \cite{Jain2023, Zheng2022, Papaioannou2022, Bohrmann2020}. Additionally, anticipatory cues—whether visual \cite{Karjanto2018}, auditory \cite{Kuiper2020}, or vibrational \cite{Li2022}—have been shown to lower MS in vehicle passengers by enabling them to predict upcoming movements \cite{Forster2020}. This effect is particularly pronounced in drivers, who experience lower MS than passengers due to their ability to anticipate motion and adjust their posture accordingly (e.g., leaning into turns) \cite{Rolnick1991, Fukuda1976, Bos2008}. In contrast, passengers often lean against centrifugal forces, misaligning their head and torso with the Gravito-Inertial Force (GIF) \cite{Wada2010}, which exacerbates MS symptoms. Studies indicate that actively leaning into turns, mimicking driver postural adjustments, significantly reduces MS symptoms in passengers \cite{Wada2012, Wada2016, Wada2018, Yusof2020}.

This phenomenon is assumed to extend to vection-induced motion perception in VR simulators and gaming environments \cite{Bos2008}. However, to the best of our knowledge, no study has explicitly examined whether individual alignment with the angles of the virtual GIF can mitigate vection-induced cybersickness. To address this gap, we conducted a pilot study to investigate whether aligning body posture with a virtual trajectory (e.g., sitting upright in straight segments and leaning into turns) reduces vection-induced cybersickness, similar to how postural alignment in vehicles reduces MS \cite{Wada2018}. By exploring this alignment-based mitigation strategy, this pilot study strives to offer new insights into cybersickness prevention and/or mitigation in VR applications in which vection is an integral feature.

\section{Methodology}


The methods for this pilot study were initially described in the preliminary research paper \textit{``MATE-AV: A VR-based training environment to teach occupants how to adopt a comfort-oriented postural control in a vehicle"}~\cite{Croucher2025}. This research paper focused on the effectiveness of the VR-based postural training mechanisms in the context of passengers in automated vehicles and mitigating MS. However, in this paper, we present the analysis for the secondary objective of that project, namely, mitigating cybersickness during simulated driving experiences.

\subsection{Participants}
Sixteen individuals (7 women and 9 men; age range = 21 - 35, mean = 24.50, SD = 3.20) took part in this experiment. They were recruited from Delft University of Technology via the university channels and were assigned equally to the two experimental groups: a VR-training group and a control group. Inclusion criteria required normal or corrected-to-normal vision, no self-reported vestibular disorders, and no history of severe motion-sickness susceptibility. Before participation, each person completed the Visually Induced Motion Sickness Susceptibility Questionnaire (VIMSSQ)~\cite{Keshavarz2023} to assess predisposition to cybersickness. All participants provided written informed consent, and ethical approval was obtained from the Human Research Ethics Council of Delft University of Technology (Delft, Netherlands, application number: 3598).

\subsection{Experimental Design and Procedure}
\begin{figure} [t]
    \centering
    \includegraphics[width=1\linewidth]{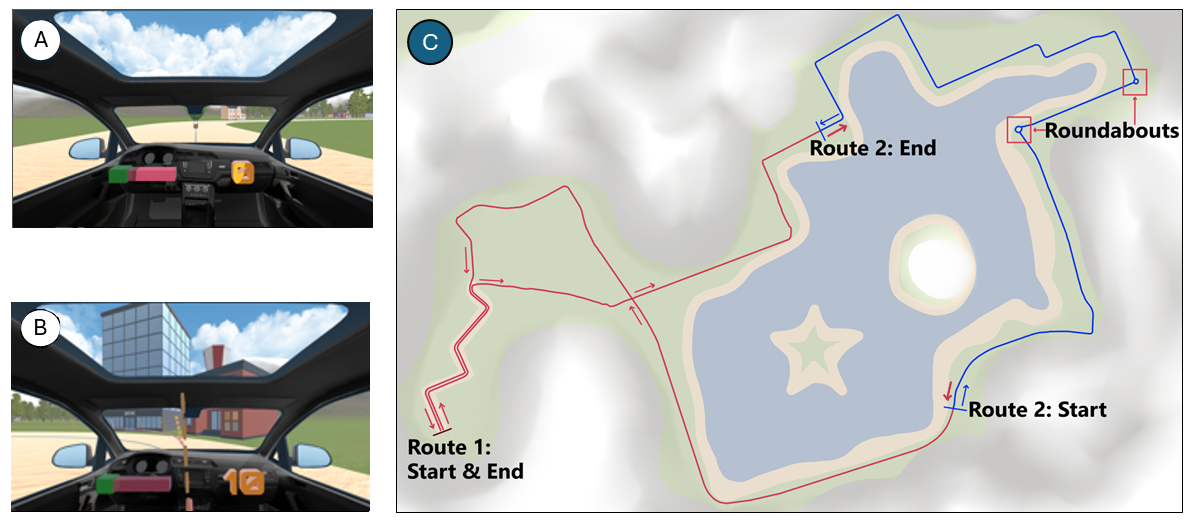}
    \caption{Images A and B are screenshots of MATE-AV and correspond to the hot air balloons and the coin alignment task, respectively. Image C shows a map view of the two routes in MATE-AV \cite{Croucher2025}. The condition route (route 1) is the entire length of the route (red and blue sections), whilst the pre-condition and post-condition routes (route 2) are only the blue section.}
    \label{fig:MATE_AV}
\end{figure}

A mixed-design approach was employed to examine whether an active postural alignment task could influence vection-induced cybersickness over time. Upon arrival, each participant completed the VIMSSQ questionnaire~\cite{Keshavarz2023}, provided informed consent, and then received instructions regarding the VR equipment and the experimental tasks. Participants were then randomly assigned to either the VR-training group or the control group. For all participants, the experiment involved three sequential VR driving routes (experienced as a passenger), referred to as the \textit{pre-condition}, \textit{condition}, and \textit{post-condition} routes. 
The \textit{pre-condition route} served as a baseline and lasted approximately seven minutes, during which participants completed an initial VR drive to establish baseline cybersickness symptoms and postural behaviours. They then completed the \textit{condition route}, which lasted about 24 minutes and included up to three short pauses, each lasting no more than five minutes. The condition route contained the key experimental manipulation (presence or absence of the postural alignment task). During this route, those in the VR-training group were instructed to align their trunk and head movements with visual-auditory concurrent cues that were presented as coins (figure \ref{fig:MATE_AV}b) as part of the active postural alignment task, and thus aligning with the virtual GIF in the corners and the straight segments. Whilst those in the control group simply observed hot air balloons (figure \ref{fig:MATE_AV}a) without making any deliberate postural adjustments. The post-condition route was the same as the pre-condition route and lasted about seven minutes, which allowed for the assessment of any changes in cybersickness symptoms and postural behaviours following the condition route. Between each of the three routes, participants removed the head-mounted display (HMD) for a two-minute break, intended to minimise residual cybersickness~\cite{Kourtesis2024} and provide a consistent transition between the route sessions.
During all VR routes, participants self-reported their Fast Motion Scale (FMS) ratings~\cite{Keshavarz2011} every minute; these ordinal scores range from 0 (no motion sickness) to 20 (severe motion sickness). 

\subsection{Materials and Apparatus}
The routes in the virtual driving environment were derived from real-world driving data ~\cite{Harmankaya2024} that featured low accelerations and decelerations, inducing only low to mild motion sickness levels ($\sim$ 3 average MISC (Motion Illness Symptoms Classification scale \cite{reuten2021feelings}). Two distinct VR routes, based on this data, were used to differentiate the condition route from the pre- and post-condition routes. Throughout the experiment, participants wore a Meta Quest 3\footnote{(Meta, United States)~\cite{Quest3}} headset, which provided inside-out tracking and continuous logging of positional data (recorded continuously throughout the experiment in cm and degrees, with regards to the x, y and z position of the HMD at a sampling rate of: 100 data points per second), enabling subsequent comparison of participants’ postural alignment with the route’s angular trajectory. All participants were seated on a desk chair without armrests so that they could freely move their trunk. No additional motion cues, such as physical tilts or vibrations, were provided; participants’ only motion feedback came from the visual simulation.


\subsection{Data Characteristics and Preparation}
\begin{figure}
    \centering
    \includegraphics[width=1\linewidth]{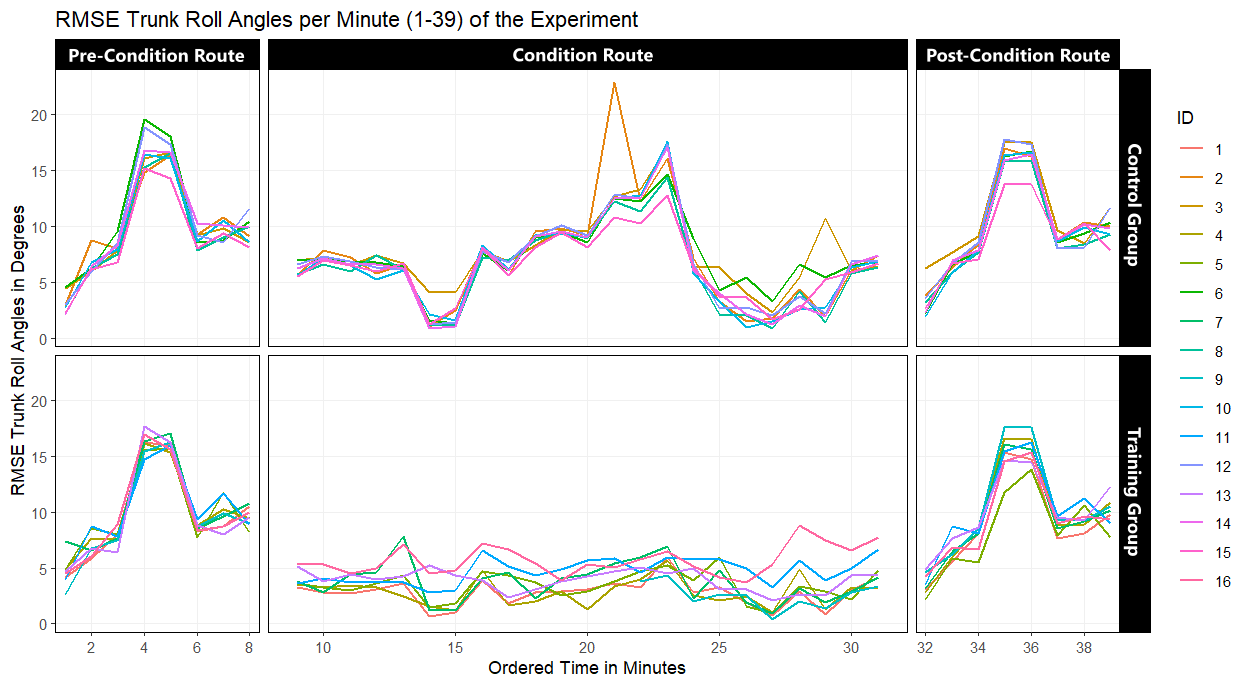}
    \caption{RMSE Trunk Roll Angles per Minute per Participant for the Entire Experiment}
    \label{fig:RMSETrunk_LineGraph}
\end{figure}

\begin{figure} [t!]
    \centering
    \includegraphics[width=1\linewidth]{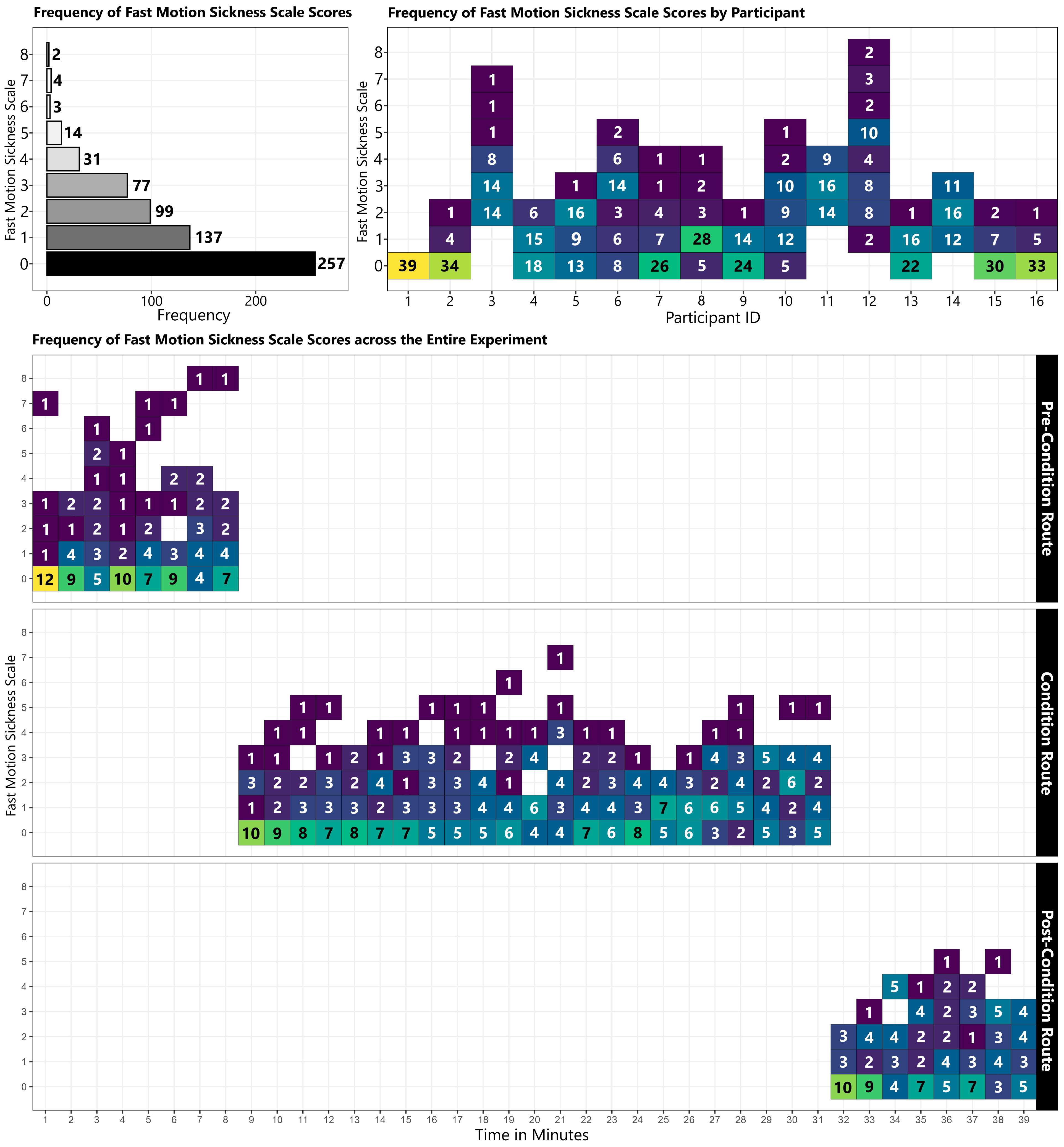}
    \caption{Total frequency of FMS scores from the entire experiment.}
    \label{fig:FMS_Frequency}
\end{figure}

All HMD positional data (x, y, z) were recorded in real-time and later converted into trunk roll angles, following the procedure detailed in~\cite{Croucher2025}. These angles were compared against the driving route angles to obtain a per-minute root mean squared error (RMSE) value (\textit{RMSE\_Trunk}), which served as an index of how closely participants aligned their posture with the implied motion of the VR vehicle (figure \ref{fig:RMSETrunk_LineGraph}). Table~\ref{tab:InitialDataDistribution} provides an overview of the dataset, including the number of positional data points and FMS scores. Although the FMS scale extends to a maximum of 20, the highest score reported in this dataset was 8 (figure \ref{fig:FMS_Frequency}), which may reflect the relatively mild driving profile~\cite{Harmankaya2024} and the use of low-sickening motion data. 
With regards, the participant's susceptibility to cybersickness (visually induced motion sickness) recorded at the start of the experiment using the VIMSSQ questionnaire (0 = no susceptibility to visually induced MS symptoms and 165 = high susceptibility to visually induced MS symptoms) \cite{Keshavarz2023}. The VIMSSQ values in this dataset range from 6.00 to 79.70 (mean = 31.73 $\pm$  19.15).
One participant (ID: 9) had incomplete condition-route data, beginning only after approximately 762.79 seconds, limiting the extent of within-subject comparisons for this individual. 

\begin{table} [h]
    \centering
        \caption{Total Number of Data-points for all 16 Participants in the data set}
        \resizebox{\columnwidth}{!}{%
    \begin{tabular}{|lr|r|r|r|}
    \hline
         \multicolumn{2}{|l|}{\textbf{Data}} & \textbf{Pre-Condition Route} & \textbf{Condition Route} & \textbf{Post-Condition Route}\\
    \hline
        \multicolumn{2}{|l|}{\textbf{Trunk Roll Angles}} & 800,016 data-points & 2,392,537* data-points & 800,016 data-points \\
         & RMSE\_Trunk & 128 data-points & 355* data-points & 128 data-points\\
    \hline
         \multicolumn{2}{|l|}{\textbf{FMS Scores}} & 128 data-points & 368 data-points & 128 data-points\\
    \hline
    \multicolumn{5}{|l|}{*Participant 9 is missing some data corresponding to positional angles in the condition route.}\\
    \hline
    \end{tabular}%
}
    \label{tab:InitialDataDistribution}
\end{table}

\subsection{Statistical Analysis}
The primary goal of the analysis was to investigate whether individuals who more closely matched the VR route’s angular profile (i.e., exhibited lower \textit{RMSE\_Trunk}) reported fewer cybersickness symptoms. To explore this relationship, a cumulative link mixed model (CLMM) was adopted~\cite{Taylor2022,Reinhard2017}, as the FMS scores are ordinal (ranging from 0 to 8 in the present dataset).

The model was fitted in R (Version~4.4.1) using the \texttt{clmm} function within the \texttt{ordinal} package (version 2023.12-4.1)~\cite{Christensen2023}, applying a logit link function, flexible thresholds, and Laplace approximation. The dependent variable was the FMS score, while the fixed effects included the per-minute \textit{RMSE\_Trunk}, time in minutes (1 to 39 across all routes), the categorical variable for the three routes (pre-condition, condition, post-condition) coded as 1 - 3 to represent VR exposure, group assignment (VR-training or control), and VIMSSQ scores. A random intercept was specified for each participant to account for within-subject variability. The model is formalised as follows:

\begin{equation}
\label{Equation1}
\begin{split}
\text{FMS Score}_i \sim \, & \text{RMSE\_Trunk}_i + \text{Time in Minutes}_i + \text{Route}_i \\
& + \text{Group}_i + \text{VIMSSQ}_i + (1 \mid \text{ID}_i). 
\end{split}
\end{equation}

Odds ratios for statistically significant predictors were computed by exponentiating the respective fixed-effect estimates (\(\exp(\beta)\)). Following Chen et al.~\cite{Chen2010}, odds ratios below 1.68 were deemed small effects. Two versions of the model were estimated, one including the partial data of Participant~9 and one excluding them outright. Data wrangling and visualisation were carried out using R packages such as \texttt{tidyverse}~\cite{Tidyverse2019}. Our supplementary material contains the results of our statistical analysis ensuring transparency in data processing and analysis.
\section{Results}
The results of our two CLMMs with and without Participant 9 indicate the same significance, and the model without Participant 9 only marginally improves in AIC (with Participant 9, AIC = 1407.34; without Participant 9, AIC = 1361.86). Therefore, we choose to report the model with Participant 9 included; please refer to the supplementary material for both results and code for the statistical analysis. 
Our model (with Participant 9) indicates that the random intercept for participants had a variance of 3.865 ($\pm$ 1.966), indicating high variability in participants' FMS scores. The following sections will report the results of our predictors (fixed effects): trunk roll alignment, total time, route order, group and cybersickness susceptibility. 

\subsection{Trunk Roll Alignment and the Relationship on FMS Scores}
To investigate the relationship between trunk roll alignment with the angles of the routes in MATE-AV (RMSE Trunk Roll) on self-reported cybersickness (FMS scores), we conducted a CLMM (Equation 1).
The results suggest that RMSE for trunk troll alignment (after adjusting for the other predictors - Equation 1) is a statistically significant predictor of self-reported cybersickness (FMS scores) ($\beta$ = 0.050, Standard Error (SE) = 0.020, z = 2.506, p = .012). Suggesting that for each unit increase on the log-odds scale regarding RMSE trunk roll values (decrease in trunk roll alignment with the angles of a virtual route), the odds of self-reporting a higher cybersickness score (FMS) increase approximately by 5\% (Odds ratio = 1.05), suggesting a gentle increase as a predictor variable. 

\subsection{Time in Minutes and the Relationship on FMS Scores}
To investigate the relationship between total time in minutes (ordered 1 to 39 minutes) and self-reported cybersickness (FMS scores), we conducted a CLMM (Equation 1).
The results suggest that total time in minutes (after adjusting for the other predictors - Equation 1) is a statistically significant predictor of self-reported cybersickness (FMS scores) ($\beta$ = 0.102, SE = 0.016, z = 6.227, p = \textless.001). Suggesting that for each unit increase on the log-odds scale regarding time in minutes (minute increments), the odds of self-reporting a higher cybersickness score (FMS) increases approximately by 11\% (Odds ratio = 1.11), suggesting another gentle but marginally higher increase as a predictor variable than RMSE trunk roll alignment. 

\subsection{VR Exposure (Routes) and the Relationship on FMS Scores}
To investigate the relationship between continued VR exposure (the three routes, pre-condition, condition, and post-condition routes were coded as 1 to 3 to represent continued VR exposure) and self-reported cybersickness (FMS scores), we conducted a CLMM (Equation 1).
The results suggest that continued exposure to VR across the three routes (after adjusting for the other predictors - Equation 1) is a statistically significant predictor of self-reported cybersickness (FMS scores) ($\beta$ =  -1.372, SE = 0.283, z = -4.842, p = \textless.001). Suggesting that for each unit increase on the log-odds scale regarding continued VR exposure (via the ordered routes, 1 - 3), the odds of self-reporting a higher cybersickness score (FMS) decrease approximately by 75\% (Odds ratio = 0.25).

\subsection{Groups and the Relationship on FMS Scores}
To investigate the relationship between the two groups (control and VR-training) and self-reported cybersickness (FMS scores), we conducted a CLMM (Equation 1).
The results suggest that the groups (after adjusting for the other predictors - Equation 1) were not statistically significant predictors of self-reported cybersickness (FMS scores) (\textit{training group:} $\beta$ = -0.993, SE = 1.114, z = -0.891, p = .373). Suggesting that whether the participants received the VR-training or not, this was not a predictor of reporting higher or lower FMS scores in the experiment. 

\subsection{Participants' Susceptibility to Cybersickness and the Relationship on FMS Scores}
To investigate the relationship between participants' susceptibility to cybersickness (VIMSSQ) and self-reported cybersickness scores (FMS scores), we conducted a CLMM (Equation 1).
The results suggest that the participants' susceptibility to cybersickness (VIMSSQ) (after adjusting for the other predictors - Equation 1) is a statistically significant predictor of self-reported cybersickness scores (FMS scores) ($\beta$ = 0.078, SE = 0.030, z = 2.598, p = .009). Suggesting that for each unit increase on the log-odds scale regarding cybersickness susceptibility, the odds of self-reporting a higher cybersickness score (FMS) increases approximately by 8\% (Odds ratio = 1.08), suggesting another gentle increase as a predictor variable. 

\subsection{Model Threshold Coefficients}
In table \ref{tab:CLMM_coefficients}, we report the threshold cutoff coefficients on the latent scale from our CLMM. These threshold cut-off coefficients indicate that the lower FMS scores are more likely to be reported, as noted in the smaller threshold coefficients, especially if we consider that an increase in FMS score from 0 to 1 is only 0.760, indicating a slight increase required in the latent variable thus the predictor values do not need to be as high. However, according to the threshold cut-off coefficients, higher predictor values are necessary to reach FMS scores above 5. Still, because the difference between the thresholds decreases from an FMS score of 5 (\textless 1.2), the increase in FMS scores is predicted to progress more quickly if reached. These threshold cut-off coefficients align with previous research on FMS scores and analysis with CLMMs in that higher FMS scores need higher predictor values but, once reached, progress quickly \cite{Reinhard2017}.   

\begin{table} [!h]
    \caption{Threshold cutoff coefficients from the CLMM model (with Participant 9)}
    \centering
    \begin{tabular}{|l|r|r|r|r|}
        \hline
        Threshold Coefficient & Estimate $\beta$ &  Difference to Prior Threshold & SE & z\\
        \hline
         $0|1$ & 0.760 & N/A & 1.423 & 0.534\\
         $1|2$ & 2.724 & 1.964 & 1.427 & 1.909\\
         $2|3$ & 4.445 & 1.721 & 1.438 & 3.091\\
         $3|4$ & 6.069 & 1.624 & 1.451 & 4.182\\
         $4|5$ & 7.250 & 1.181 & 1.467 & 4.943\\
         $5|6$ & 8.393 & 1.143 & 1.495 & 5.615\\
         $6|7$ & 8.839 & 0.446 & 1.514 & 5.838\\
         $7|8$ & 9.998 & 1.159 & 1.622 & 6.166\\
        \hline
    \end{tabular}
    \label{tab:CLMM_coefficients}
\end{table}

\section{Discussion}
This pilot study aimed to investigate whether actively aligning one’s trunk with a virtual route—sitting upright during straight segments and leaning into turns—can mitigate vection-induced cybersickness. Using a CLMM, we found that each unit of trunk roll misalignment was associated with a 5\% increase in the odds of reporting cybersickness, while each additional minute spent in VR increased those odds by 11\%. In contrast, repeated exposures led to a 75\% decrease in cybersickness risk, suggesting desensitisation or habituation effects, while participants with a higher baseline susceptibility were 8\% more likely to experience symptoms. These findings align with MS research, in which active trunk and head movements are known to reduce symptoms \cite{Rolnick1991, Fukuda1976, Wada2018}. Taken together, this pilot study's findings highlight the potential of active postural alignment as a complementary measure to hardware- and software-based cybersickness mitigation techniques.

Specifically, structured session design, incorporating progressive exposure and frequent breaks, appears especially beneficial for long-duration VR. In addition, user-specific susceptibility underscores the need for personalised approaches. Practitioners and researchers can improve both comfort and immersion by aligning the posture of the trunk, managing the exposure time, and tailoring the experiences to individual needs. These insights therefore open new avenues for designing VR simulations, training applications, and entertainment scenarios that minimise cybersickness while maximising user engagement. In what follows, we discuss how our results fit into the broader literature on cybersickness mitigation techniques, consider the role of individual differences and session design, and note key limitations that inform future investigations towards optimising the design of VR applications encompassing vection.

\subsection{Trunk Roll Alignment and Cybersickness}
Consistent with the notion that vection-induced discomfort can be alleviated by anticipatory signals \cite{Karjanto2018, Bos2008}, our results indicate that for every unit increase in trunk roll misalignment relative to the virtual trajectory (GIF), the chances of reporting a higher cybersickness score increased by 5\%. This aligns with previous studies that emphasise postural control in VR \cite{Palmisano2020, Palmisano2022, Park2025}, which have focused mainly on head positioning and head movement. By demonstrating that trunk alignment also matters, our findings broaden current perspectives on how best to mitigate the sensory conflict inherent in vection-rich scenarios \cite{Reason1978, Riccio1991}.

Importantly, many existing cybersickness countermeasures rely on hardware or software optimisations \cite{Kundu2021, Groth2024, Weech2020}, such as blurred motion or reduced rendering latency. Although these may be effective, they do not address user-driven responses to perceived motion. Our results suggest that deliberately matching one's posture to visual motion cues can reduce sensory conflict and, therefore, reduce the risk of cybersickness. Future VR applications - especially simulators or immersive entertainment experiences - can harness this insight by providing explicit postural cues (that is, sensory information such as visual, auditory, olfactory and/or tactile) \cite{Phlmann2024, Jung2021, Grassini2021}, prompting users to lean with turns and straighten up in straight segments. Any such system should carefully consider both head and torso movements for maximum efficacy.

\subsection{Time, VR Exposure, and Cybersickness}
A second key finding relates to exposure duration and repeated sessions. Each additional minute in VR increased the odds of reporting higher cybersickness scores by 11\%, mirroring previous evidence that vection-induced discomfort can accumulate over time \cite{Davis2014, Risi2019}. At the same time, repeated exposure led to a 75\% decrease in those odds, suggesting that habituation or desensitisation processes strongly moderate the severity of symptoms \cite{Reinhard2017, Zhang2015}. Taken together, these observations imply that while prolonged VR sessions might escalate the risk of cybersickness, structured repetition can help users build tolerance.

This has practical implications for designing longer-term VR interventions. For example, training or rehabilitation protocols may prioritise shorter sessions, with intermittent breaks or cooling phases, over a single extended immersion. This approach is reminiscent of established MS habituation protocols \cite{Zhang2015}, in which incremental controlled exposures elicit adaptation. Researchers and developers could integrate these insights to reduce dropout rates and enhance user comfort, allowing more sustained engagement with VR applications.

\subsection{Individual Susceptibility}
Our analysis also confirms that individual susceptibility is a significant contributor to cybersickness severity. Higher VIMSSQ scores were associated with an 8\% increased likelihood of experiencing more pronounced symptoms, aligning with the existing literature linking susceptibility to motion sickness to discomfort in VR \cite{Nooij2021, Tian2022, Papaefthymiou2024,Kourtesis2024}. This underscores the difficulty of applying a uniform mitigation strategy in diverse user populations.

Instead, personalised approaches may be warranted. Users with higher susceptibility might benefit from more frequent breaks, lower intensity exposures, or personalised postural guidance - potentially aided by real-time feedback on trunk and head alignment. The integration of these user-specific strategies within immersive systems can help ensure broader accessibility, paralleling the calls for adaptive solutions in VR \cite{Kourtesis2024b, Stanney2020}. By acknowledging and accommodating variability in susceptibility, VR developers can further minimise cybersickness and improve the overall user experience.

\subsection{Limitations and Future Directions}
Although our preliminary findings are promising, this exploratory analysis has limitations. First, 41\% of the FMS scores were reported as zero and the highest score reached was eight (on a scale of 0-20), indicating that participants generally experienced mild levels of cybersickness. This probably reflects the relatively gentle motion profile of our real-world driving data \cite{Harmankaya2024}, which provoked low MISC levels. Second, the sample size was small and the high variability in susceptibility scores constitutes a challenge in interpreting certain effects.  

Future research could address these issues by employing more provocative motion trajectories or by systematically recruiting participants across a broader range of cybersickness susceptibility \cite{Tian2022}. It would also be informative to evaluate whether trunk alignment remains effective under more extreme vection conditions (e.g. VR roller coasters). Finally, longitudinal studies could track how postural adaptation, repeated exposures, and individual susceptibility interact over long timescales.

\section{Conclusion}
Our findings postulate that postural alignment in vection-rich virtual reality scenarios appears to be a promising avenue to mitigate cybersickness. In particular, actively matching one’s trunk posture to the virtual trajectory, such as leaning into turns, can help reduce visual-vestibular mismatches that commonly lead to cybersickness. In addition, individuals vary in their susceptibility to cybersickness, suggesting that adaptive or personalised approaches may further improve comfort and accessibility in VR experiences. Finally, exposure duration and repeated sessions also influence how users adapt over time. Taken together, these findings underscore the potential value of integrating postural strategies into broader mitigation practices, offering insights that may benefit a wide range of VR applications, from training to entertainment.
\section*{Supplemental Materials}
\label{sec:supplemental_materials}
Placeholder

\section*{Acknowledgements}
We want to thank Floris Pauwels for collecting this data and also our Hi-Drive partners for their support and feedback. 

\section*{Funding}
The research that led to these results was funded by the European Union's Horizon 2020 Research and Innovation Programme under grant agreement No 101006664 (Hi-Drive).

\section*{Author Contributions}
\textbf{C.C:} Conceptualisation, Methodology, Formal Analysis, Writing - Original Draft, Writing - Review \& Editing, Visualisation \textbf{P.K:} Writing - Original Draft, Writing - Review \& Editing, Supervision \textbf{G.P:} Writing - Review \& Editing, Supervision, Project administration, Funding acquisition



 \bibliographystyle{elsarticle-num} 
 \bibliography{Manuscript}

\end{document}